# ONE MEMS DESIGN TOOL WITH MAXIMAL SIX DESIGN FLOWS

*Honglong Chang, Jinghui Xu, Jianbing Xie, Chengliang Zhang, Zijian Yan, Weizheng Yuan*

Micro and Nano Electromechanical System Laboratory, Northwestern Polytechnical University, China, email: changhl@nwpu.edu.cn, tel: +86-29-88495102

**ABSTRACT**

This paper presents one MEMS design tool with total six design flows, which makes it possible that the MEMS designers are able to choose the most suitable design flow for their specific devices. The design tool is divided into three levels and interconnected by six interfaces. The three levels are lumped-element model based system level, finite element analysis based device level and process level, which covers nearly all modeling and simulation functions for MEMS design. The six interfaces are proposed to automatically transmit the design data between every two levels, thus the maximal six design flows could be realized. The interfaces take the netlist, solid model and layout as the data inlet and outlet for the system, device and process level respectively. The realization of these interfaces are presented and verified by design examples, which also proves that the enough flexibility in the design flow can really increase the design efficiency.

## 1. INTRODUCTION

MEMS design tools help the designers to model and simulate various MEMS devices, which have been expected to play an important role in the commercialization of MEMS as the EDA tool did in the success of microelectronics industry. Consequently the development of the MEMS CAD is directed toward one EDA-like software suite to a large extent. The system level behavior simulation and mask layout design tools are all borrowed from EDA directly. And it seems like that the top-down design concept, which dominates the mainstream in microelectronics, has become one impressing feature for almost all current commercial MEMS CAD software suites. It addresses the hierarchical synthesis and optimization from the system to the final mask layout. [1-5] However, the diversities of MEMS devices are more and more challenging for this structured design method for MEMS.

By now various MEMS devices such as gyroscopes, pressure sensors and micro mirrors with different principles, structures and processes have been invented, which makes it nearly impossible to use such one unified design flow to maximize the design efficiency of every MEMS device. For example, for the capacitive gyroscopes the top down design flow starting from the system level then directly to the process level could increase the efficiency as expectation, but for the capacitive pressure sensors the flow doesn't work properly because the system level behavior modeling based on lumped-element model can not solve such space-continuous models properly. Therefore to establish one design tool that could provide the most suitable design flow according to the specific device's need is one practical and effective way to accelerate the MEMS design process. Actually many commercial MEMS CAD software have begun to include various design entries at different levels to increase the flexibility of the design flow.

This paper will focus on the techniques about how to increase the flexibility of choosing design flows. And one MEMS design tool with maximal six design flows under the current popular three-level structure will be established, based on which the designers can start the design process from any level and finish it as a whole closed loop.

## 2. FRAMEWORK OF THE MEMS CAD

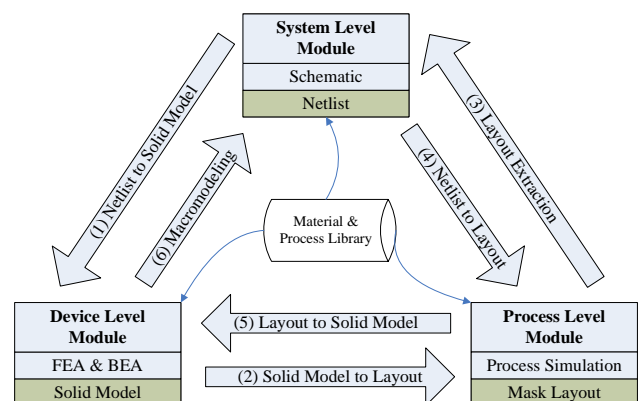

Figure 1: Framework of the proposed MEMS design tool with six data transmitting interfaces.

As shown in Fig.1 the proposed MEMS design tool consists of three levels and six interfaces. Same as other commercial MEMS CAD software, the three levels include most modeling and simulation functions such as parameterized component library to support the behavior modeling and simulation, FEA simulation for the device level and layout design with process simulation in the process level, etc. Therefore the three levels are the foundation of the MEMS design tool and the order of appearance for these levels will decide the design flow of one MEMS device.

To reach the maximal flexibility of the design flow, total six interfaces are proposed to connect the different levels by transmitting the design data. The number of the interfaces reaches maximum under this popular three-level MEMS CAD structure. Thus the three levels and six interfaces form into one closed design triangle, the designers are able to begin one design flow from any level, then to any other levels ending it as a closed loop finally. The popular "top-down" and "bottom-up" design flows are also incorporated in these six design flows. As an example, the gyroscope can use the flow as "System Level→ Interface (1) →Device Level→ Interface (2)→ Process Level", while for the pressure sensors the flow such as "Device Level→ Interface (6) →System Level" is maybe the most proper one.

Based on this framework, establishing the necessary material library, process simulation module and parameterized layout library etc, the complete MEMS CAD software could be set up.

### 3. INTERFACES REALIZATION

The six interfaces are defined as data transmitting between the three levels, therefore choosing the proper data representation form for each level is the first step to realize the interfaces. As the system level module in the paper takes the analog hardware description language (AHDL) based simulator, the design data will be rendered through the netlist file finally. For the device level the 3D solid model is one proper form for meshing and the further FEA simulation. While for the process level the 2D layout is the widely accepted representation form. Consequently in this paper the netlist, solid model and layout are treated as the best data representation form for the system level, device level and process level separately. With the proper representation for each level the data transmitting is converted into a computer programming problem. However the interface from the device level to the system level is not simply the data format changing, it is usually called macromodeling and has been studied by many former researchers. [3, 4] The realization of these interfaces will be discussed in detail.

**3.1. Netlist to Layout & Solid model**

The two interfaces from the system level is the key to realize the top down design concept, which have been implemented by different method. [5-7] In this paper a mechanical schematic is used for capturing the design intent of a MEMS device in the system level. As shown in Fig.6, the schematic is composed of various parameterizable components such as masses, beams, and interdigitated capacitive comb fingers etc. All of the information about the device such as geometry size, location and process information are saved in the netlist file. Therefore to extract proper information from the netlist file is the key to realize the two interfaces from the system level.

The extracting process is divided into four steps, i.e. reading files, information retrieval and storage, information processing and standard layout & solid model generation. The first three steps are shown in Fig.2, among which the key step to generate the solid model and layout is shown in Fig.3 in detail.

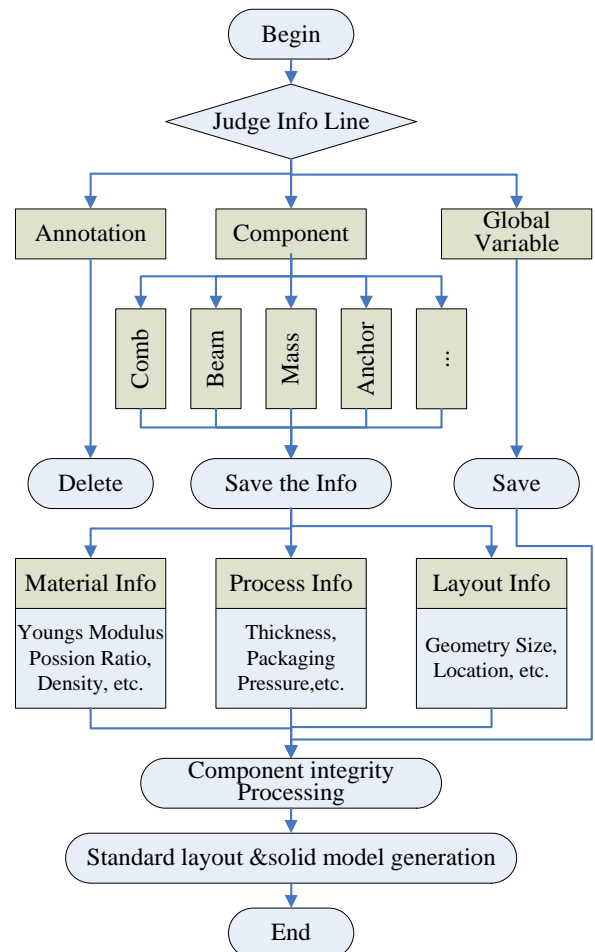

Figure 2: The algorithm of solid model and layout generation from netlist.

**3.2. Solid Model to Layout**

The direct layout generation from the optimized solid model has been long expected and studied by many researchers. [1,5,8,9]. The interface could be thought of as one of the most difficult steps in the top-down synthesis flow and expected to greatly increase the design efficiency. Especially for those engineers with mechanical background, the visual 3D solid model design is more intuitive than the system level schematic building or the 2D layout editing.

the *.SAT file as the saving medium, and the layout is represented by the *.CIF file. Extracting the geometry, topology and property information from the SAT file could be used for generation of the CIF file. The corresponding algorithm is shown in Fig.4.

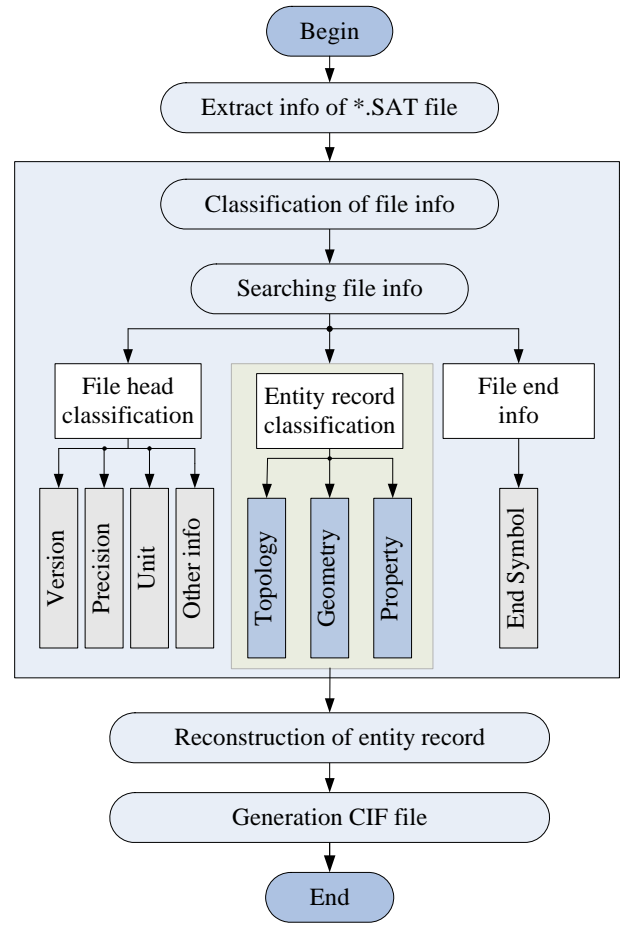

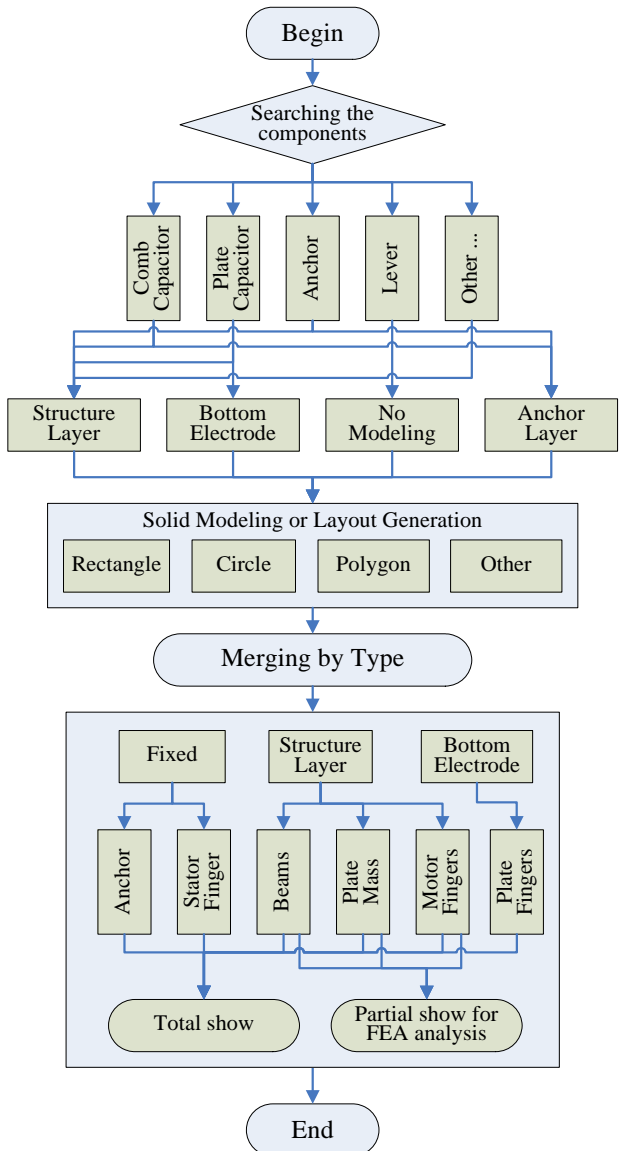

Figure 3: The detailed algorithm of solid model generation (layout generation is similar).

Here we presented one novel method via the transfer of standard file format to realize the interface firstly. In this method the solid model of the MEMS devices all take

Figure 4: The algorithm flow of solid model to layout.

### 3.3. Macromodeling

The interface from device level to the system level is not simply the change of data format but the change of degree of freedom for the device models and usually called as macromodeling. A macromodel is a low-order behavioral representation of a device. Usually it has the attracting attributes such as correct and explicit energy conservation and dissipation behavior, covering both quasi-static and dynamical behavior, expressible in a simple-to-use form such as an equation or a network analogy or a small set of coupled ordinary differential equations, easy to connect to system-level simulator, etc.

In this paper, the macromodeling process is realized through the operation on FEA result files. The realization process is as shown in Fig.5, where the model order reduction is based on the Arnoldi algorithm. [10,11]. The Arnoldi method projects a original system onto a subspace spanned by $\{v_0, Av_0, A^2v_0, \ldots, A^{k-1}v_0\}$ using the modified Gram-Schmidt process.[11] It operates on a $N$ by $N$ matrix $A$, a vector $B$ of size $N$ and a predefined reduced system's order $q$. Since the transfer function of the reduced linear system approximates that of the original linearized system but often with $q<<N$, the expected model order reduction is achieved.

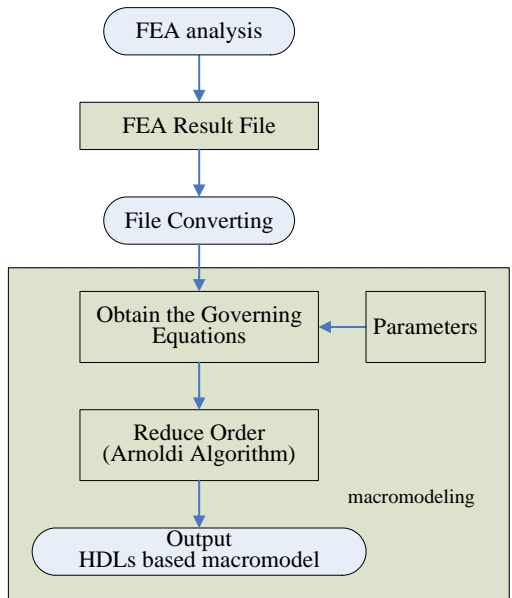

Figure 5: The algorithm flow of macromodeling based on the FEA analysis.

### 3.4. Other Interfaces

Among the six interfaces the one from layout to solid model is commonly implemented in current commercial MEMS CAD software; here the similar techniques were taken and the details of the technique will be ignored due to the page limit.

The interface from the process level to the system level is different from the macromodeling. It recognizes geometry pattern of the typical components such as masses, comb drives and beams from the layout and correspond to the components in the system level. [12, 13] To some extent this interface has big limitations due to the complexity of the layout and varieties of the process.

### 4. DESIGN EXAMPLES

To demonstrate the advantages of this MEMS design tool owning maximal flexibility of choosing flows, we presented two typical MEMS devices' design examples. The examples covered five interfaces except the interface from layout to system level.

### 4.1. System→Solid Model→Layout

The first device is one capacitive gyroscope with comparatively simple structure topology. [14] And the chosen design flow is system level first and then device and process levels sequentially.

The mechanical schematic for the MEMS gyroscope captured using Saber tool is shown in Fig. 6. [15] It consists of sixteen 3D beams, four rigid masses, six linear comb capacitors, four bias comb capacitors and eight anchors. Based on this schematic various analyses could be carried out. [14] After simulation the netlist file was used to generate the solid model of the gyroscope for the further FEA simulation (Fig.7). To save the numerical simulation time the comb fingers could be cut in the solid model editor. After the optimized design in the device level, the mask layout was generated from the solid model file. The mask layout could be edited further according to the specific processes for the final fabrication. (Fig.8)

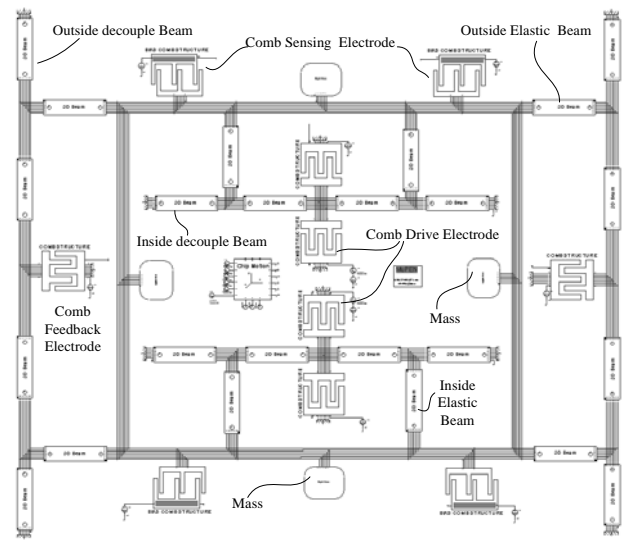

Figure 6: The mechanical schematic for gyroscope.

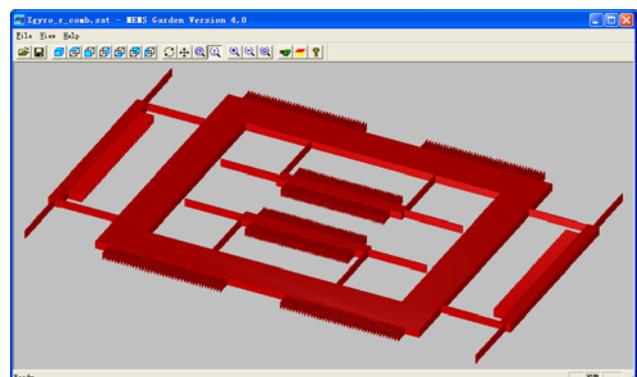

Figure 7: The solid model of the movable structure of the gyroscope, the comb fingers could be cut off for the further FEA simulation.

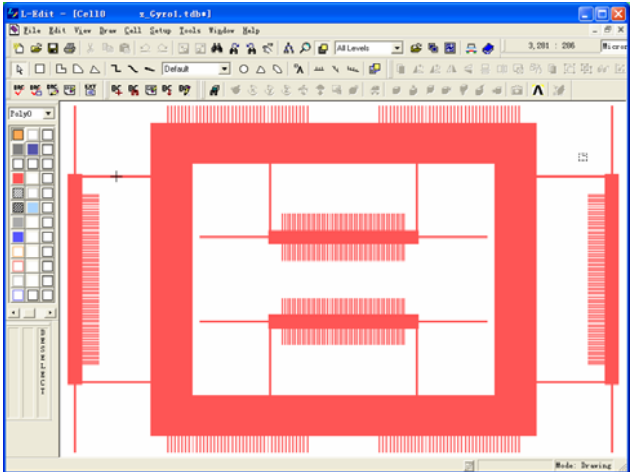

Figure 8: Layout generated from solid model and input into the usual layout editor. [15]

### 4.2. Layout→Solid Model→System

The above flow is only suitable for those devices that the components can be found in the reusable libraries. Furthermore if the number of the components is too big then the system level simulation will cost much time and the convergence will be impossible. The design flow from the process level and then device level and system sequentially could solve this problem. Herein the other capacitive z-axis accelerometer with more complex suspension is chosen as an example to demonstrate the design flow.

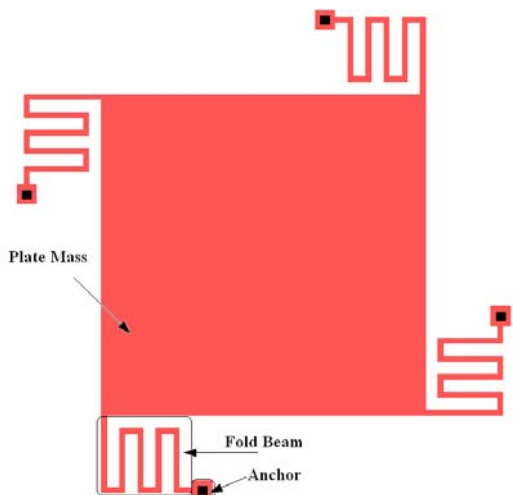

Figure 9: Layout of the z-axis accelerometer.

As shown in Fig.9 the z-axis accelerometer has a very complex topology, the beams of which are ten folded. To model this device using the components is somewhat time-consuming and the simulation time will be very long. Therefore the designers can firstly draw the layout in the process level then generate the solid model from the layout for FEA analysis by the corresponding interface. (Fig.10) After the FEA simulation the result file is used to extract macromodel of the complex beams through the macromodeling interface. (Fig.11) Then the system level model of the accelerometer could be established, based on which the accelerometer could co-simulate with the interface circuit. (Fig.12) The transient analysis results with one pulse and one sinusoidal signal input was shown in Fig.13. Compared with FEA analysis the macromodel-based system level simulation could cost much less time with approximate accuracy.

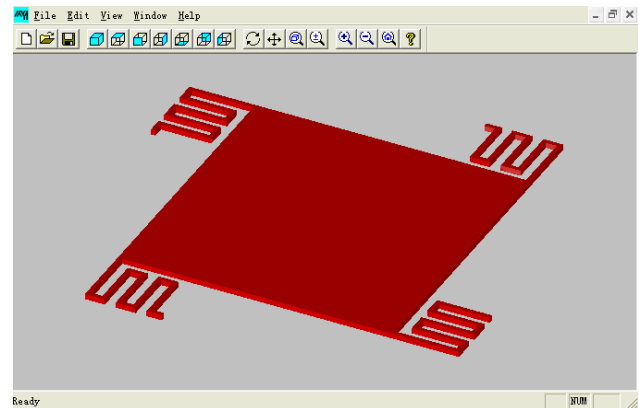

Figure 10: Solid model generated from the layout.

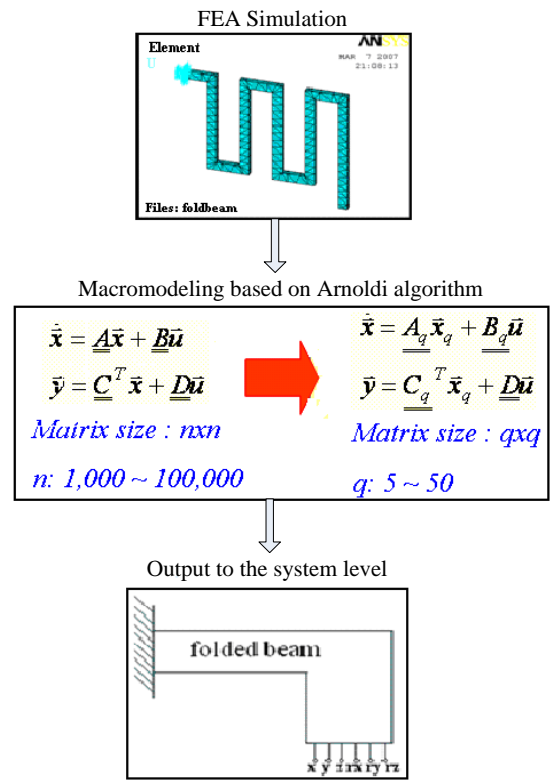

Figure 11: Macro-modeling process of the folded beams

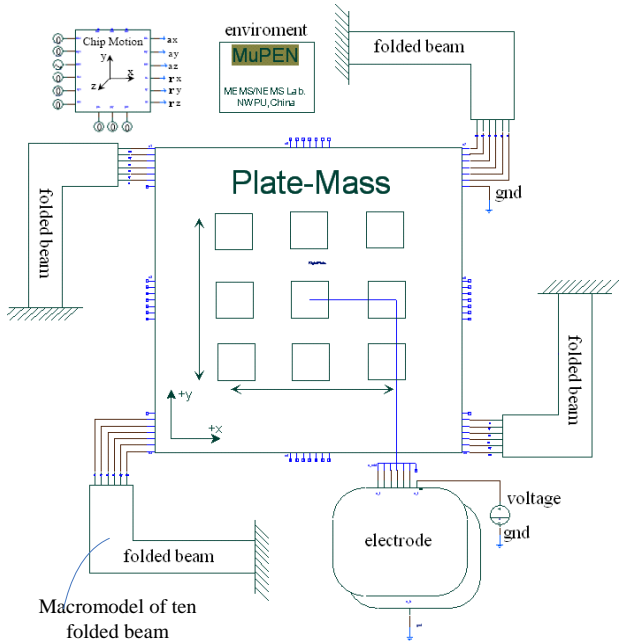

Figure 12: Mechanical schematic for the z-axis accelerometer.

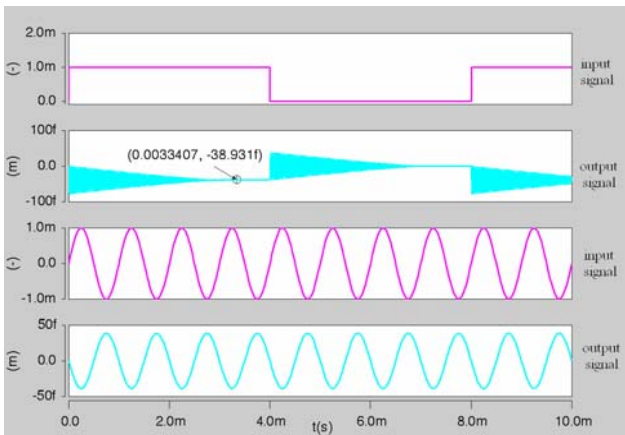

Figure 13: MEMS-IC co-simulation results of the z-axis accelerometer.

## 5. CONCLUSIONS

From the above description it can be found that the six interfaces really could provide the maximal flexibility in the choosing design flows. MEMS designers could begin the design and simulation process from his familiar entry according to the specific devices. Thus the design efficiency could be improved.

On the other hand, the realization of most interfaces takes the standard data format as the media of data transmitting, which could provide the best compatibility for different software. Thus the users will not to buy new software package if they have already own some similar software.

## 6. REFERENCES

[1] E.K. Antonsson, "Structured design methods for MEMS final report", *A Workshop sponsored by the National Science Foundation*, Pasadena, CA, 1995

[2] G.K. Fedder, "Top-down design of MEMS". *Procs. of the International Conference on Modeling and Simulation of Microsystems*, San Diego, CA, USA, 2000, pp.7-10

[3] S.D.Senturia, "CAD Challenges for Microsensors, Microactuators, and Microsystems", *Proceedings of the IEEE*, Vol. 86, no. 8, 1998, pp. 1611-1626

[4] M.H. Zaman, S.F. Bart, & J.R. Gilbert, "An Environment for design and modeling Electro-Mechanical micro-systems", *Journal of modeling and simulation of Microsystems,* Vol.1, no.1, 1999, pp. 65-76

[5] W. Tang, "Overview of Microelectromechanical systems and design processes". *Procs. Of Design Automation Conference*, CA, USA, 1997, pp.670-673

[6] G. Lorenz, S. Remy Chevreuse, "System and method for three-dimensional visualization and post-processing of a system model", *US Patent 2005/0125750*

[7] G. Lorenz, S. Remy Chevreuse, & C.J. Kennedy, "System and method for automatic mesh generation from a system-level MEMS design", *US Patent 2005/006630*

[8] V. Ananthakrishnan, "Part-to-Art: Basis for a Systematic Geometry Design Tool for Surface Micromachined MEMS", *PHD thesis of University of Toledo*, Dec, 2000.

[9] M. Lang, D. David, & M. Glesner, "Automatic Transfer of Parametric FEM Models into CAD-Layout Formats for Top-down Design of Microsystems". *Procs. Of European Design & Test Conferences*, Paris, France, 1997, pp.200-204

[10] Lv Xianglian, Yuan Weizheng, Yan Zijian, & Li Jun. "A Macro-model Extraction method based on Finite Element Analysis". *Procs. Of the Asia-Pacific Conference of Transducers and Micro-Nano Technology*, Singapore, 2006

[11] G.H. Golub, C.F.V. Loan, "Matrix Computations", *the Johns Hopkins University Press*, 1996

[12] B. Baidya, S.K. Gupta, & T. Mukerhejee, "MEMS component extraction", *Procs. of International Conference on Modeling and Simulation of Microsystems,* 1999, pp.143-146

[13] B. Baidya, S.K. Gupta, & T. Mukherjee, "An Extraction-Based Verification Methodology for MEMS", *Journal of Microelectromechanical Systems,* Vol.11, no.1, 2002, pp.2-11

[14] Honglong Chang, Weizheng Yuan, Jianbing Xie, Qinghua Jiang, Chengliang Zhang, "One Mechanically Decoupled Z-axis Gyroscope", *Proceedings of IEEE NEMS*, Zhuhai, China, 2006, pp.373-376

[15] Saber, Ansys, L-Edit is the trademark of Synopsys Inc, Ansys Inc, and Tanner Inc separately.